\documentstyle[amstex,righttag,intlim]{amsart}%
\newenvironment{enumeraterm}{%
\begin{enumerate}}{\end{enumerate}%
}
\newcommand{\CinfO}{C^\infty_0}                                 % ()
\newcommand{\Cinf}{C^\infty}                                    % ()
\newcommand{\supp}{\operatorname{supp}}                         % Traeger
\newcommand{\identity}{\mbox{${\bf 1}$}}                        % Identity
\newcommand{\WF}{\mbox{\rm WF}}                                 % Wave-
\newcommand{\Rn}{{{\Bbb R}^n}}                                  % R^n
\newcommand{\cO}{{\cal O}}                                      % \cal O
\newcommand{\cC}{{\cal C}}                                      % Cauchyfl.
\newcommand{\cCb}{{{\cal C}_b}}                                 % Umgebung

\newtheorem{Thm}{Theorem}
\newtheorem{Cor}{Corollary}
\newtheorem{Prop}{Proposition}
\theoremstyle{definition}
\newtheorem{Dfn}{Definition}
\newtheorem{exmp}{Example}
\theoremstyle{remark}
\newtheorem{rem}{Remark} 
\newtheorem{ack}{Acknowledgement} 
\begin{document}
{\tt \noindent DESY 94-161 \hfill ISSN 0418-9833\\
     hep-th/9410027 \hfill \\}
\title{New examples for Wightman fields on a manifold}
\author[M. K\"ohler]{M. K\"ohler$\dag$\\
        {II.\ Institut f\"ur Theoretische Physik, Universit\"at
          Hamburg\\
        Luruper Chaussee 149, D--22761 Hamburg\\
        Germany}}
\thanks{$\dag$ Supported by the DFG}
\thanks{E-mail: mkoehler@@x4u.desy.de}
%\address{II. Institut f\"ur
%  Theoretische Physik\\
%  Universit\"at Hamburg\\
%  Luruper Chaussee 149\\
%  D--22761 Hamburg\\
%  Germany}
%\email{mkoehler@@x4u2.desy.de}
\date{September 21, 1994}
\dedicatory{{\rm PACS numbers: 1110, 0460, 0230}}
\maketitle
\begin{abstract}
  The product of two free scalar fields on a manifold is shown to be
  a well defined operator valued distribution on the GNS Hilbert
  space of a globally Hadamard product state. Viewed as a new field
  all n-point distributions exist, giving a new example for a Wightman
  field on a manifold.
\end{abstract}
%
%\tableofcontents
%
\section{Introduction}
Quantum field theory on curved spacetimes describes quantum fields
propagating in a classical curved background. One of the main
difficulties with such systems comes from the absence of Poincar\'e
symmetry. On flat spacetime this symmetry fixes the vacuum states or
equivalently allows a preferred class of representations of the
canonical commutation relations (CCR) to be picked out. In the general
case no analogous selection criterion exists. Since field theories
have infinitely many degrees of freedom, different states may lead to
unitarily inequivalent representations of the CCR\@. To deal with this
fact the algebraic approach to quantum field theory by Haag and
Kastler~\cite{haag:64} seems appropriate. In this setting, roughly
speaking, one describes the theory by a `net of local algebras' which
encodes the fields and observables.  The positive linear functionals
with unit norm on this net are called states. They describe the
preparation of the system.  However not all states are believed to be
physical states. Nevertheless every state fixes --via the GNS
construction-- a Hilbert space, a `vacuum vector' and a representation
of the algebras, thus links the algebraic approach to the usual
Hilbert space setting.  Yet one might end up with unitarily
inequivalent representations. Therefore it is not sufficient to
construct the net; it is also necessary to characterize these physical
states. One approach to this characterization are the `scaling limit
criterion' and the `principle of local definiteness' introduced by
Haag, Narnhofer and Stein~\cite{haagnarnhofer:84} and further
investigated by Fredenhagen and Haag in~\cite{fredenhagen:87}. Both
characterizations are designed for states on general quantum field
theoretical models on arbitrary spacetimes. On the other hand, a class
of states which is believed to be physical for the free, linear models
are the quasifree Hadamard states: The singularity structure of their
two-point distributions fulfils the `Hadamard condition', e.\ g.\ it
is fixed by the underlying geometry.  (For a review
see~\cite{kay:91,Verch:94} or the book of
Fulling~\cite{Fulling:aspects_of_qft} and the references therein)
However, besides from being a global condition, this condition has the
disadvantage of being restricted to linear (free) models.

Radzikowski showed recently in~\cite{Radzikowski:92} that the global
Hadamard condition can equivalently be formulated locally in terms of
wave front sets. Moreover, using wave front sets, he formulated a
spectrum condition for quantum fields on a manifold, which might be
useful as a new selection criterion for physical states. Unfortunately
it can be shown that the n-point distributions of the free scalar
field on a general spacetime violate his condition for $n>2$. We
propose a slightly modified definition of a wave front set spectrum
condition (WFSSC) for two-point distributions and give a nontrivial
example of a Wightman field on a general spacetime satisfying this
modified WFSSC\@.  Starting from Hadamard states of two free,
noninteracting Klein-Gordon fields, it is shown that their product is
a new Wightman field on the spacetime together with a product state
whose two-point distribution satisfies the new spectrum condition.
Furthermore it will become clear that our method can be applied to
other construction schemes initially based on Minkowski spacetime as
well.

The organization of the work is as follows: After the Introduction we
state the main result of this paper. For the convenience of the reader
we continue by quoting some notions and results from the theory of
pseudo\-differential operators concerning distributions on manifolds;
they are well known in the mathematical literature, but apparently
were seldom used in physics.  The notion of the wave front set of a
distribution on a manifold is defined next and we relate this concept
to Hadamard states.  Our version of Radzikowski's spectrum condition
follows.
Finally the main Proposition is proved.
\section{The product of two free scalar fields on a curved spacetime}
We consider two free, noninteracting scalar fields $A$ and $B$
propagating on a spacetime which we assume to be a four dimensional,
orientable, time orientable, globally hyperbolic Lorentz manifold with
metric signature $(+,-,-,-)$.  The
classical Lagrangian of our model is given by
\[
{\cal L}= \frac{1}{2} \left({A}_{;\mu} {A}^{;\mu} + ( \xi R - m^2)
{A}^2 + {B}_{;\mu} {B}^{;\mu} + ( \xi R - m^2) {B}^2 \right).
\]
$R$ is the scalar curvature and $\xi$ describes an additional coupling
to the geometry.  The equation of motion for the two fields obviously
decouples giving the Klein-Gordon equation for both $A$ and $B$.  We
define a new field $K$ to be the pointwise product of the basic fields
$A$ and $B$ respectively:
\begin{equation}
  K(x) := A(x) \cdot B (x)
  \label{eq:product}
\end{equation}
Consider two quasifree `globally Hadamard' states
$\sideset{^A}{}{\omega}$ and $\sideset{^B}{}{\omega}$ for the two
scalar fields $A$ and $B$.  Their tensor product $\omega \equiv
\sideset{^A}{}{\omega} \otimes \sideset{^B}{}{\omega}$ defines a state
for our model; it is called a {\em globally Hadamard product state}.
The quantized fields $A$ and $B$ are mutually commuting operator
valued distributions on the GNS Hilbertspace of $\omega$.  Note that
the r.h.s.\ of Eqn.~(\ref{eq:product}) denotes the distributional
product of $A$ and $B$ after quantization and such a product is not
necessarily well defined a priori. The main result of this note is
\begin{Prop}\label{prop:strom}
  $K$ is well defined and gives a new Wightman field on the GNS Hilbertspace
  $(H,\Omega,A,B)$ of a quasifree globally Hadamard product state
  $\omega$; the two-point distribution of $K$ in this product state
  $\omega$ fulfils the wave front set spectrum condition
  (Definition~\ref{dfn:wfspectrum} below).
\end{Prop}
For the proof of this proposition we need some properties of the wave
front set of distributions, which are included in this work for the
convenience of the reader. For further details and the proofs which
are omitted, we refer the reader to the original literature
(\cite{Hoermander:71,Hoermander:72}) or to the monographs of Taylor
and Reed \& Simon~\cite{Taylor:81,reed:MethodsII}.

The theory of wave front sets was developed in the seventies by
H\"ormander together with Duistermaat for their studies of
pseudodifferential operators and differential equations on
manifolds~\cite{Hoermander:71,Hoermander:72}.  Wave front sets ($\WF$)
are refinements of the notion of the singular support (sing supp) of a
distribution. One main reason for using them in favor of the sing supp
is that they provide a simple characterization for the existence of
products of distributions and eliminate the difference between local
and global results. It is interesting that Duistermaat and H\"ormander
even mention relations between their `microlocal analysis' which they
used to study solutions of the Klein-Gordon equation on a manifold and
quantum field theory.  It seems however that their results did not
find their way into the physical literature prior to the work of
Radzikowski~\cite{Radzikowski:92}.
\begin{Dfn} \label{def:sigma_z}
  Let $v \in {\cal D}' (\Rn)$ be a distribution on ${\Bbb R}^n$.  The
  set $\Sigma_z (v)$ is the complement in ${\Bbb R}^n \setminus \{ 0
  \}$ of the set of all nonzero $\xi \in \Rn $ for which there is a
  smooth function $\phi$ with compact support not vanishing at $z$ and
  a conic neighborhood $C_\xi$ of $\xi$ such that for all $N \in \Bbb
  Z$ there exists a constant $C_N$ such that for all $\xi' \in
  C_{\xi}$
   \[ (1 + \left| \xi' \right| )^N \left| \widehat{\phi u}(\xi') \right|
                              \leq C_N \]
The hat $\hat{~}$ denotes Fourier transformation.
\end{Dfn}
\begin{Dfn}\label{def:wavefrontset}
The wave front set of $v$ is defined by
\begin{equation}
\WF(v) = \{ (z,\theta) \in T^*\Rn \setminus \{ 0 \}| \quad
                  \theta \in \Sigma_z (v)\}
\label{wavefront}
\end{equation}
\end{Dfn}
All points $<x,\xi> \in T^*\Rn$ which are {\em not\/} in the
wave front set are called {\em regular directed points}.
\begin{rem} \label{remark}

  \begin{enumerate}
  \item $\WF{}(v)$ is a closed subset of $T^*\Rn \setminus \{0\}$
    since each regular directed point $<x,\xi> \not\in \WF(v)$ has by
    definition an open neighborhood in $T^*\Rn \setminus \{0\}$
    consisting of regular directed points, too.
  \item\label{rem:wfsmooth}
    The wave front set of a smooth function is the empty set
  \item For $v \in {\cal D}'(\Rn)$ with wave front set
    \WF$(v)$, the projection of the wave front set to the base point
    gives the singular support of $v$.
  \item\label{rem:wfinklusion} For all smooth functions $\tilde{\phi}$
    with compact support $\WF(\tilde{\phi}v) \subset \WF(v)$.
  \item\label{rem:addition}
    For two distributions $v,w \in {\cal D}'(\Rn)$ with wave front sets
    \WF$(v)$ and \WF$(w)$ respectively, the wave front set of $(v + w)
    \in {\cal D}' (\Rn)$ is contained in $\WF(v) \cup \WF(w)$
  \end{enumerate}
\end{rem}
In order to extend this definition to manifolds we use the behavior of
$\WF(v)$ under diffeomorphisms of $\Rn$ to $\Rn$
(Theorem~\ref{thm:TransWF} below). Although
Definition~\ref{def:sigma_z} contains a Fourier transform, it turns
out that the elements of $\WF(v)$ transform as elements of $T^*\Rn$,
the cotangential bundle of $\Rn$. It is therefore possible to define
the wave front set of a distribution on a manifold via charts.
\begin{Thm}[Theorem~IX.44 and Problem~75 in~\cite{reed:MethodsII}]
  \label{thm:TransWF}
  Let $v \in {\cal D}'(\Rn)$ be a distribution on $\Rn$, $\chi$ be a
  diffeomorphism of $\Rn$ to $\Rn$ and let $v \circ \chi$ be the
  distribution
  \[ v \circ \chi (f) := v( g^{-1} ( f \circ \chi^{-1})) \]
  where $g$ is the determinant of the Jacobian matrix $d\chi$. Define
  $\chi_* : \Rn \times (\Rn \setminus \{0\}) \rightarrow
  \Rn \times (\Rn \setminus \{0\}) $ by
  \[ \chi_* <x,\xi> \; = \; <\chi(x), d\chi^*(\xi)> \]
  where $d\chi^*$ is the adjoint of $d\chi$ with respect to the
  Euclidean inner product on $\Rn$. Then
  \[ \WF(v \circ \chi) = \chi_* \left( \WF(v) \right) \]
\end{Thm}
\begin{Dfn}\label{dfn:wfmanifold}
  Let $ u \in {\cal D}'(M) $ be a distribution over some manifold $M$.
  Let $\{X_\lambda\}_{\lambda \in {\Bbb N}}$ denote an open covering
  of $M$.  Choose a compatible partition of unity
  $\{\Phi_\lambda\}_{\lambda \in {\Bbb N}}$. Without loss of
  generality we may assume that $\supp \Phi_\lambda$ is contained in a
  single coordinate patch for every $\lambda \in {\Bbb N}$. The
  corresponding charts are denoted by $ \chi_\lambda : \supp
  \Phi_\lambda \subset U \rightarrow \Rn$.  Using the same notation as
  in Theorem~\ref{thm:TransWF} above we find as a Corollary
\begin{equation}\label{wfcovering}
  {\chi_\lambda^{-1}}_* \left( \WF(\Phi_\lambda u) \right) \equiv
  \WF(\Phi_\lambda u \circ \chi_\lambda^{-1})
\end{equation}
  where $\Phi_\lambda u \circ \chi_\lambda^{-1}$ obviously is a
  distribution with compact support over $\Rn$.  We define the {\em wave
    front set\/} of $u$ by
  \[  \WF(u) = \bigcup_\lambda \WF(\Phi_\lambda u) \]
\end{Dfn}
\begin{rem}
  This is a {\em local\/} definition. It follows that the
  following results which are stated in~\cite{reed:MethodsII} for
  distributions on $\Rn$, are valid in the generic case, too.
\end{rem}
A useful application of wave front sets is the definition of products
of distributions.  Wave front sets provide a simple characterization
for the existence of such products, which furthermore can be extended
to manifolds.  The following definition of a product and its relation
to wave front sets can be found for example in Reed \&
Simon~\cite[p.90--97]{reed:MethodsII}.
\begin{Dfn}
  Let $v,w \in {\cal D}'(\Rn)$. The distribution $T\in {\cal D}'(\Rn)$
  is the {\em product\/} of $v$ and $w$ if and only if for all $x \in
  \Rn$ there exists a smooth function $f$ not vanishing at $x$ such
  that for all $k \in \Rn$
  \begin{equation}
    \widehat{f^2T}(k) = (2\pi)^{-n/2} \int_{\Rn} \widehat{fv}(l)
    \widehat{fw}(k-l) d^nl
    \label{eq:productRn}
  \end{equation}
  where the integral is absolutely convergent.
\end{Dfn}
\begin{rem}
  \par
  \begin{itemize}
  \item The product is well defined, since if such a $T$ exists, it is
    unique: \\
    Let $g \in {\cal D}(\Rn)$ then
    \[
    \widehat{gf^2T} = (2\pi)^{-n/2} \widehat{gfv} * \widehat{fw} =
    (2\pi)^{-n/2} \widehat{fv} * \widehat{gfw},
    \]
    since the change of variables is legitimate due to the assumption
    of absolute convergence of (\ref{eq:productRn}). Now suppose $T_1$
    and $T_2$ both fulfil (\ref{eq:productRn}), e.g.\ for all $x \in
    \Rn$ there exist functions $f$ and $g$ not vanishing at $x$ such
    that $\widehat{f^2T_1}=\widehat{fv} * \widehat{fw}$ and
    $\widehat{g^2T_2}=\widehat{gv} * \widehat{gw}$.
    We conclude $\widehat{f^2g^2T_1} = \widehat{f^2g^2T_2}$, so
    $(T_1-T_2)$ vanishes near $x$ for all $x$, that is it is zero.
  \item The product of two distributions with disjoint support is
    zero.
  \item Since this is a local definition, we can extend it immediately
    to manifolds, using charts (See~\cite{Hoermander:71}).
  \end{itemize}
\end{rem}
The following Theorem gives the relation between the wave front sets
of two distributions and the existence of their product.
\begin{Thm}[Theorem IX.54 of \cite{reed:MethodsII}]\label{Thm:IX.54}
  Let $v,w$ be two distributions on $M$ such that
  \[
  \WF(v) \oplus \WF(w) \equiv \{ <x,k_1 + k_2> | \quad <x,k_1> \in \WF(v),
  <x,k_2> \in \WF(w)  \}
  \]
  does not contain any element of the form $<x,0>$, then the product
  $v \cdot w$ exists and has wave front set
  \[
  \WF(v\cdot w) \subset \WF(v) \cup \WF(w) \cup (\WF(v) \oplus \WF(w))
  \]
\end{Thm}
\message{Section \thesection}
\section{Hadamard states}\label{sec:hadamard}
The Hadamard condition for quasifree states of scalar fields on a
manifold is believed to be a necessary condition for physical states
since the work of DeWitt and Brehme in 1960~\cite{DeWitt:60} and was
intensively studied since this time by various authors (See the
references in Fulling's book~\cite{Fulling:aspects_of_qft}).  However
only recently Kay and Wald~\cite{kay:91} gave a mathematically
rigorous definition of this condition.

In this section we recall their precise characterization of the global
Hadamard condition. We continue by calculating the wave front set of
the corresponding states using the `Propagation of Singularities
Theorem' of Duistermaat and H\"ormander (Theorem~6.1.1
of~\cite{Hoermander:72}). The reverse
relation, e.g.\ a characterization of globally Hadamard states by
their wave front sets due to Radzikowski~\cite{Radzikowski:92}, is
quoted next; it is followed by our version of his wave front set
spectrum condition.

Let $\omega_2$ be the two-point distribution of a quasifree state on
the Borchers Uhlmann algebra~\cite{borchers:62,uhlmann:62} for a
scalar field on a globally hyperbolic manifold $(M,g_{ab})$ satisfying
the Klein-Gordon equation and local commutativity.  Assume that a
preferred time orientation has been chosen on $(M,g_{ab})$ and let $T$
be a global time function increasing towards the future.  Recall the
definition of a convex normal neighborhood: An open subset $U \subset
M$ is called a convex normal neighborhood if for all points
$x_1,x_2 \in U$ there exists a unique geodesic {\em contained\/} in
$U$ connecting $x_1$ and $x_2$.  Let ${\cO} \subset M \times M$ be
an open neighborhood in $M \times M$, of the set of causally related
points $(x_1,x_2)$ such that $J^+(x_1) \cap J^-(x_2)$ and $J^+(x_2)
\cap J^-(x_1)$ are contained within a convex normal neighborhood. As
usual $J^{\pm}(x)$ denotes the causal future (past) of
the point $x$. The square of the geodesic distance, $\sigma$, is well
defined and smooth on $\cO$. For each integer $p$ and $\epsilon \geq
0$ define for $(x,x') \in \cO$ the complex valued function
\begin{multline} \label{eq:kernel}
G_\epsilon^{T,p} (x,x') \\= \frac{1}{{(2 \pi)}^2}
  \left( \frac{\Delta^{1/2} (x,x')}
         {\sigma (x,x') + 2 i \epsilon t + \epsilon^2}
        + v^{(p)}(x,x') \ln [\sigma(x,x')+2 i \epsilon t +
                         \epsilon^2]
  \right)
\end{multline}
where $t \equiv T(x) - T(x')$ and $\Delta^{1/2}$ and $v^{(p)}$ are
smooth functions uniquely determined by the geometry on $M$
($\Delta^{1/2}$ is the van Vleck Morette determinant and $v^{(p)}$ is
given by the Hadamard recursion relation up to order
$p$~\cite{DeWitt:60,castagnino_harari:84}.  The branch cut of the
logarithm is taken to lie on the negative real axis). Let $\cC$ be a
Cauchy surface and let $N$ be a neighborhood of $\cC$ with the
property: For each pair of points $x_1,x_2 \in N$ such that $x_1$ can
be reached by a causal curve emerging from $x_2$ ($x_1 \in
  J^+(x_2)$) one can find a convex normal neighborhood in $M$
containing $J^-(x_1) \cap J^+(x_2)$. $N$ is called a
causal normal neighborhood of $\cC$.  Now let $\cO '$ be a
neighborhood in $N \times N$ such that its closure is contained in
$\cO$ and choose a smooth real valued function $\chi$ on $M \times
M$ with the properties $\chi (x,y) = 0$ if $(x,y) \not\in \cO$ and
$\chi(x,y) = 1$ if
$(x,y) \in \cO '$ respectively.
\begin{Dfn}[Globally Hadamard states]\label{dfn:glob_hadam}
We say the state mentioned above is  {\em
globally Hadamard\/} iff its two-point
distribution
$\omega_2$ is such that for each integer $p$ there exists a $C^p$-
function $H^{(p)}(x,y)$ on $N \times N$ such that for all $F_1,F_2
\in C_0^\infty(N)$ we have
\begin{multline} \label{ScalarHadam}
\omega_2(F_1 \otimes F_2) \\=
                       \lim_{\epsilon \rightarrow 0}
                       \left( \quad
   \int_{N \times N} \left( \chi (x,y) G_\epsilon^{T,p}(x,y) +
        H^{(p)}(x,y) \right) F_1(x) F_2(y) d\mu_x d\mu_y
                        \right)
\end{multline}
\end{Dfn}
For a detailed discussion of this definition the reader is referred to
the work of Kay and Wald~\cite{kay:91}.
\begin{rem}
  Kay and Wald prove in~\cite{kay:91} that this Definition is
  independent of the Cauchy surface $\cC$. This in turn means that
  the Cauchy evolution preserves the Hadamard structure, e.\ g.\
  $\omega_2$ restricted to a causal normal neighborhood $N$ of a
  Cauchy surface $\cC$ already fixes $\omega_2$ throughout the whole
  spacetime.
\end{rem}
\subsection{The wave front set of a globally Hadamard state}
The result of this subsection (Corollary~\ref{Cor:wavefrontscalar}
below) can already be found in the dissertation of Radzikowski.
However the proof he gave is --as it stands-- only valid for the
rather trivial case of a flat spacetime. The proof presented here
verifies his claim in the general case.

Following Radzikowski's line of argument, consider first the free
massive Klein-Gordon field on Minkowski space in the vacuum state.
The vacuum is known to fulfil the global Hadamard condition with
$\Delta^{1/2} \equiv 1$ and $v^{(p)} \equiv \sum_{n=0}^p 2
\fracwithdelims(){m^2}{2}^{n+1} \bigl/ n!\,
(n+1)!$~(\cite{castagnino_harari:84}).  The wave front set of the
corresponding two-point distribution
\mbox{$\sideset{^{\text{Mk}}}{^m_2}{\omega}$} is also well known (See
for instance~\cite{reed:MethodsII}):
\begin{Thm}[Theorem IX.48 of~\cite{reed:MethodsII}]\label{Thm:WfMinkScal}
  The two-point distribution \mbox{$\sideset{^{\text{{\rm
          Mk}}}}{^m_2}{\omega}$} of
  the free massive scalar field on Minkowski space in the vacuum state
  has wave front set
  \begin{equation}
    \begin{split}
      & \WF(\sideset{^{\text{\rm Mk}}}{^m_2}{\omega})\\
      & = \{ (x_1,k_1),(x_2,k_2) \in {\Bbb R}^4 \times ( {\Bbb R}^4
      \setminus \{0\}) | \quad x_1 \neq x_2; \; (x_1 -x_2)^2 =0;\\
      & \phantom{ = \{ (x_1,k_1),(x_2,k_2) \in {\Bbb R}^4 \times ( {\Bbb R}^4
      \setminus \{0\}) | \quad}
      k_1 \| (x_1 -x_2); k_1+k_2 =0;\\
      & \phantom{= \{ (x_1,k_1),(x_2,k_2) \in {\Bbb R}^4 \times ( {\Bbb R}^4
      \setminus \{0\}) | \quad }
                         k_1^0 \geq 0 \} \\
      & \phantom{=\quad}
      \bigcup
      \{ (x,k_1),(x,k_2) \in {\Bbb R}^4 \times ({\Bbb R}^4\setminus
      \{0\}) | \quad k_1 + k_2 = 0; \; k_1^2=0 ; \; k_1^0 \geq 0 \}
    \label{eq:wfscalarMink}
  \end{split}
\end{equation}
\end{Thm}
For a proof see~\cite{reed:MethodsII} or use the following
representation of the Fourier transform of
\mbox{$\sideset{^{\text{\rm Mk}}}{^m_2}{\omega}$}:
\[
  \widehat{\sideset{^{\text{\rm Mk}}}{^m_2}{\omega}} = (2\pi)^{-1}
  \delta(k_1 + k_2) \Theta(k_1^0) \delta(k_1^2 - m^2)
  \in {\cal S}' ({\Bbb R}^4 \times {\Bbb R}^4)
\]
Using this representation and Definition~\ref{def:wavefrontset} it is
easy to prove the Theorem explicitly.

In order to extend this result to arbitrary manifolds, let $x \in M$
be a point on a Cauchy surface $\cC$ and let $U_x$ be a convex normal
neighborhood of $x$. Denote by $\cCb \subset \cC$ an arbitrary (non
void) open subset of $\cC$ such that the domain of causal dependence,
$D(\cCb)$, is contained in $U_x$. We are going to calculate the
wavefront set of $\omega_2$ for all base points $x_1,x_2 \in D^+(\cCb)
:= D(\cCb) \cap {J}^+ (\cCb)$ using some properties of Hadamard states
on a special smooth deformation $(\hat{M},\hat{g}_{ab})$ of our
original spacetime\footnote{The author thanks R. Verch for calling his
  attention to the deformation argument of Fulling, Narcowich and
  Wald~\cite{FullingNarcowichWald:81}. Note that the metric $g_{ab}$
  restricted to $D^+(\cCb)$ is {\em not\/} flat in general; the latter
  was assumed implicitly in Radzikowski's argument at some point,
  making the following modification of his proof necessary.}.  Finally
the Propagation of Singularities Theorem
(Theorem~\ref{Thm:prop_of_sing} below) and our knowledge that
$\omega_2$ has no singularities for space-like separated points allows
us to extend this result to the whole spacetime $(M,g_{ab})$.

Given $\cC$, $x \in \cC$ and $\cCb$ it will be shown below that
there exists a globally hyperbolic spacetime
$(\tilde{M},\tilde{g}_{ab})$ with the following properties:
\begin{enumeraterm}
  \item A neighborhood $U$ of $\cC$ in $M$ is isometrically isomorphic
    to a neighborhood $\tilde{U}$ in $\tilde{M}$ and the isometry $\phi$
    is also an isometry between $\cC$ and $\tilde{\cC}$; e.\ g.\
    $(\tilde{M},\tilde{g}_{ab})$ is a smooth deformation of
    $(M,g_{ab})$.
  \item For all points $x_1,x_2 \in D^+(\cCb)$ there exists a Cauchy
    surface $\tilde{S}$ with neighborhood $\hat{U}$ in $\tilde{M}$,
    such that the metric $\tilde{g}_{ab}$ restricted to $\hat{U}$ is
    flat (i.e.\ Minkowskian) and $D(\phi(\cCb)) \subset D(\hat{U})$.
\end{enumeraterm}
Now $\omega_2$ induces  canonically a Hadamard distribution
$\tilde{\omega}_2$ on $(\tilde{M},\tilde{g}_{ab})$, since it does so
on $\tilde{U}$: Being a Hadamard distribution on $\tilde{U}$ implies,
by the Remark following Definition~\ref{dfn:glob_hadam}, that
$\tilde{\omega}_2$ is a Hadamard distribution throughout
$(\tilde{M},\tilde{g}_{ab})$.  Furthermore, the wave front set of
$\tilde{\omega}_2$ on $\tilde{U}$ determines that of $\omega$ at $U$
and vice versa by Theorem~\ref{thm:TransWF}. On the other hand, using
the Propagation of Singularities Theorem and the smoothness of
$\tilde{\omega}_2$ for space-like separated points, the wave front set
of $\tilde{\omega}_2$ at $\tilde{x}_1 = \phi(x_1)$ and $\tilde{x}_2 =
\phi(x_2)$ is already fixed by the wave front set of that distribution
at all points in $\hat{U}$ (See below and note that $\tilde{x}_1$ and
$\tilde{x}_2 \in D(\hat{U})$). To obtain the wave front set of
$\omega_2$ at $x_1$ and $x_2$ it is therefore sufficient to calculate
the wave front set of
$\tilde{\omega}_2$ for all points in $\hat{U}$.
%%%%%%%%%%%%%%%%%%%%%%%%%%%%%%%%%%%%%%%%%%

The following computations are performed entirely in the flat part of
$\tilde{M}$ and were partly sketched in~\cite{Radzikowski:92}. To
keep the notation simple we omit the $\tilde{\phantom{x}}$ in what
follows. Note first that Eqn.~(\ref{ScalarHadam}) is valid in
$(\tilde{M},\tilde{g}_{ab})$, too. Choose a point $x \in S \equiv
\hat{\cC}(t_1)$ and a convex normal neighborhood $U_x \subset
\hat{U}$ of $x$.  Using normal coordinates and an adapted time
function $T$ the distribution $G^{T,p}_\epsilon$, when restricted to
$U_x \times U_x$ is
\begin{multline*}
    G^p:=\\
    \lim_{\epsilon\rightarrow 0}\quad \chi G^p_\epsilon
    =
    \lim_{\epsilon\rightarrow 0}\quad
    \frac{\Delta^{1/2}(x_1,x_2)}{-(x_1-x_2)^2 + 2i\epsilon
      (x_1^0-x_2^0) +\epsilon^2}\\
    + v^{(p)}(x_1,x_2) \ln(-(x_1-x_2)^2 + 2i\epsilon
      (x_1^0-x_2^0) +\epsilon^2)
\end{multline*}
The metric $g_{ab}$ is flat on $U_x$ and hence $2 \sigma(x,y) \equiv
-(x_1-x_2)^2$ in these coordinates. Furthermore the Van Vleck Morette
determinant $\Delta^{1/2}$ and $v^{(p)}$ are identical to the
corresponding functions on Minkowski space, since only {\em local\/}
properties of the underlying geometry enter into the Hadamard
recursion relations.  We conclude that $\WF(\sideset{^{\text{\rm
      Mk}}}{^0_2}{\omega})=\WF(G^p)$ in normal coordinates on $U'_x
\times U'_x$. Using Definition~\ref{dfn:wfmanifold} this result can be
pulled back to the
manifold:
\begin{multline*}
  \WF(G^p)
  = \{ (x_1,k_1),(x_2,k_2) \in (T^*M \times
  T^*M) \setminus \{0\} | \\
   \quad
  (x_1,k_1) \sim (x_2,-k_2); \quad k_1^0 \geq 0 \}
  \quad \text{on $U'_x \times U'_x$}
\end{multline*}
where $(x_1,k_1) \sim (x_2,k_2)$ means (i) $x_1$ and $x_2$ can be
joined by a null geodesic $\gamma$, (ii) $k_1 (= k_{1\nu})$ is a
cotangent vector such that $k_1^\mu=k_{1\nu} g^{\mu\nu}$ is tangential
to $\gamma$, (iii) The parallel transport of $k_1$ along $\gamma$
yields $k_2$ {\em or\/} (i) $x_1=x_2$, (ii) $k_1^2=0$ and (iii)
$k_1=k_2$. Now $H^{(p)} \in C^p(M\times M)$ which implies for
$\phi \in \CinfO(M)$ and $\phi(x) \neq 0$
\[
(( {\phi\otimes\phi)H^{(p)})}\sphat (k_1,k_2) \leq C_p(1+|k|)^{-p}
\]
for a constant $C_p$ and all $k=(k_1,k_2)$. Since $\omega_2 = G^p +
H^{(p)}$ for all $p$ we conclude that the wave front set of $\omega_2$
restricted to $U_x' \times U_x'$ is
\begin{multline} \label{eq:wfscalarUx}
\WF(\omega_2) = \{ (x_1,k_1),(x_2,k_2) \in (T^*M \times T^*M)
\setminus \{0\} | \\
\quad
(x_1,k_1) \sim (x_2,- k_2); \;\; k_1^0 \geq 0 \} \quad
           \text{on $U_x' \times U_x'$}
\end{multline}
As a Corollary to the following Theorem of H\"ormander this extends to
all of $\tilde{M}$.
\begin{Thm}[Corollary to Theorem 6.1.1 of \cite{Hoermander:72}]
  \label{Thm:prop_of_sing}
  Let $P= \identity \otimes (\Box + \nabla_\mu V^\mu \linebreak[2] + b)$ be a
  pseudodifferential
  operator on a globally hyperbolic manifold $(M,g_{ab})$. $\Box$
  denotes the D'Alembert operator and $V^\mu$ and $b$ are a smooth
  vector field and smooth function on $M$ respectively. The principal
  symbol of $P$ is denoted by $p$ and is given here by
  \begin{align*}
    p: (T^*M  \times T^*M) \setminus \{ 0\}
    & \rightarrow {\Bbb R} \\
    (x_1,k_1;x_2,k_2) & \mapsto g^{\mu\nu}(x_2) k_{2\mu} k_{2\nu}
  \end{align*}
  If $u \in {\cal D}'(M\times M)$ is a weak solution of $Pu=0$ with wave front
  set $\WF(u)$, it follows that \\
  a) $\WF(u) \subseteq p^{-1}(0)$ and\\
  b) $\WF(u)$ is invariant under the (Hamiltonian) vector field $H_p$
  given by
  \[
   H_p
   = \sum_{i=1}^{2n} \frac{\partial p(x,k)}{\partial x_i}
  \frac{\partial}{\partial k_i} - \frac{\partial p(x,k)}{\partial k_i}
  \frac{\partial}{\partial x_i}
  \]
  in local coordinates.
\end{Thm}
\begin{Dfn}
  The bicharacteristic strips of $P$ are the curves on the submanifold
  $p^{-1}(0) \subset (T^*M \times T^*M) \setminus \{ 0
  \}$
  which are generated by $H_p$.
\end{Dfn}
Note that b) means: If $(x_1,k_1;x_2,k_2)$ and $(x'_1,k'_1;x'_2,k'_2)$ are on
the same bicharacteristic strip, denoted by $(x_1,k_1;x_2,k_2) \approx
(x'_1,k'_1;x'_2,k'_2)$ and $(x_1,k_1,x_2,k_2) \in \WF(u)$, then
$(x'_1,k'_1;x'_2,k'_2) \in \WF(u)$.
For $P=\identity \otimes (\Box + \nabla_\mu V^\mu + b)$ one finds (See
Proposition~2.8 in~\cite{Radzikowski:92})
\[
(x_1,k_1;x_2,k_2) \approx (x'_1,k'_1;x'_2,k'_2)
       \Leftrightarrow
(x_1,k_1) = (x'_1,k'_1) \quad \text{and}\quad (x_2,k_2) \sim (x'_2,k'_2)
\]
\begin{Cor}[See Theorem~2.6 in~\cite{Radzikowski:92}]
  \label{Cor:wavefrontscalar}
  The two-point distribution $\omega_2$ of a free massive Klein-Gordon
  field on a globally hyperbolic spacetime in a globally Ha\-da\-mard
  state has wave front set
\begin{multline}\label{eq:wfhadamscalar}
  \WF(\omega_2) = \{ (x_1,k_1),(x_2,-k_2) \in (T^*M \times
  T^*M ) \setminus \{0\} | \\
  \quad
  (x_1,k_1) \sim (x_2,k_2); \quad k_1^0 \geq 0 \}
\end{multline}
\end{Cor}
\begin{pf}
  Since $\omega_2$ is a bisolution of the Klein-Gordon operator, we
  can apply Theorem~\ref{Thm:prop_of_sing} with $P_1= \identity{}
  \otimes (\Box + m^2)$ or $P_2 = (\Box + m^2) \otimes \identity$.
  In the first case we conclude by a) that
  \[
  \WF(\omega_2) \subseteq \{ (x_1,k_1;x_2,k_2) \in (T^*M \times  T^*M)
  \setminus \{ 0 \} | \quad k_2^2 =0 \}
  \]
  and --using b)--
  \begin{multline}\label{eq:propk_2}
  (x_1,k_1; x_2,k_2) \in \WF(\omega_2)
  \wedge (x_2,k_2) \sim (x'_2,k'_2)
  \Rightarrow
    (x_1,k_1; x'_2,k'_2) \in \WF(\omega_2)
  \end{multline}
  Obviously the second case leads to
  \[
  \WF(\omega_2) \subseteq \{ (x_1,k_1;x_2,k_2) \in (T^*M \times  T^*M)
  \setminus \{ 0 \} | \quad k_1^2 =0 \}
  \]
  and
  \begin{multline}\label{eq:propk_1}
  (x_1,k_1; x_2,k_2) \in \WF(\omega_2)
  \wedge  (x_1,k_1) \sim (x'_1,k'_1)\\
  \Rightarrow
    (x'_1,k'_1; x_2,k_2) \in \WF(\omega_2)
  \end{multline}
  To decide whether $(x_1,k_1;x_2,k_2)$ with $k_1^2 =k_2^2 =0$ is in
  the wave front set of $\omega_2$, one checks first whether $x_1$ and
  $x_2$ are space-like separated. If this is true, then
  $(x_1,k_1;x_2,k_2) \not \in \WF(\omega_2)$, since the projection of
  the wave front set of a distribution to the base point gives the
  singular support of that distribution; but $\omega_2$ is assumed to
  have no singularities at space-like separated points. If they are
  causally related, we use Eqn.~(\ref{eq:propk_2})
  and~(\ref{eq:propk_1}) to propagate $(x_i,k_i)$ along the null
  geodesic $\gamma_i$ with tangent vector $k_i$ at $x_i$ to the Cauchy
  surface $\cC$. Note that due to the global hyperbolicity of $M$
  this is always possible. If the points on $\cal C$ do not coincide,
  we have $(x_1,k_1;x_2,k_2) \not \in \WF(\omega_2)$, since different
  points on a Cauchy surface are space-like separated. Should they
  coincide at say $x \in \cal C$, we check whether this particular
  combination $(x,k_1;x,k_2)$ is in the wave front set using
  Eqn.~(\ref{eq:wfscalarUx}).
\end{pf}
%%%%%%%%%%%%%%%%%%%%%%%%%%%%%%%%%%%%%%%%%%%%

To finish the argument, we have to construct a spacetime
$(\tilde{M},\tilde{g}_{ab})$ that satisfies the properties (i) and
(ii). This can be done by using the methods of Appendix~C
in~\cite{FullingNarcowichWald:81} (See also~\cite{Verch:93}). Let
$D(\cCb) \subset V \subset N$ be a causal normal neighborhood of
$\cC$, such that $\cC$ is also a Cauchy surface for $V$. $V$ is, by
the normal exponential map $\phi$ of $\cC$, diffeomorphic to an open
neighborhood $\tilde{V} \subset {\Bbb R} \times \cC$ of $\tilde{\cC} =
\{ 0 \} \times \cC = \phi(\cC)$. Using normal coordinates
$(t,\mbox{{\bf x}})$ around some point in $\tilde{\cC}$, the metric
$\phi_*g$ on $\tilde{V}$ takes the form
\[ dt^2 - h_{ij}(t,\mbox{{\bf x}}) d\mbox{x}^i d\mbox{x}^j \]
Choose $t_1 < t_2 <t_3 <0$ such that
\[ \overline{ \hat{\cC}(t_i) \cap J^-(\tilde{\cCb})} \subset
     \phi(U_x) \qquad i= 1,2,3,
\]
where $\hat{\cC}(t) := \{ t \} \times \cC$, $t \in {\Bbb R}$ and
$\tilde{\cCb} := \phi(\cCb)$ (As a consequence of the global
hyperbolicity and due to the assumption $D(\cCb) \subset U_x$ such a
choice is always possible). Recall that $\phi(U_x)$ can be covered
with a single coordinate patch by assumption. Next choose a
neighborhood $\tilde{U}$ of $\tilde{\cC}$ such that
$\overline{\tilde{U}} \subset \tilde{V} \cap \mbox{{\rm int}} \left(
J^+ (\hat{\cC}(t_3)) \right)$. Due to the fact that $t_1 < t_2$ there
exists a neighborhood $\hat{U}$ of $\overline{ J^- ( \tilde{\cCb} )
  \cap \hat{\cC}(t_1) }$ in $\tilde{M}$ such that (i) $\overline{
  \hat{U}} \subset \phi(U_x)$ and (ii) $\overline{\hat{U}} \cap J^+
(\hat{\cC}(t_2)) = \emptyset $. Let $f\in \Cinf({\Bbb R}\times \cC,
{\Bbb R})$ be a smooth function $0 \leq f \leq 1$, $f\equiv 0$ on
$\tilde{U}$ and $f\equiv 1$ outside the closure of $\tilde{V}$ or in
the past of $J^-(\hat{\cC}(t_2))$. Let $\tilde{h}$ be a complete
Riemannian metric on $\cC$ being flat on the spatial component of
$\hat{U} \cap (\hat{\cC}(t_3))$. Note that the existence of such a
$\tilde{h}$ follows from the fact that $\overline{ \hat{U}}$ can be
covered by a single coordinate patch (i.e.\ $\phi(U_x)$). Let
$\beta\in\Cinf({\Bbb R}\times \cC, {\Bbb R}^+)$ be a function equal to
one on $\tilde{U}$ and on $(-\infty,t_2)$. Define a Lorentzian metric
$\tilde{g}_{ab}$ on ${\Bbb R}\times \cC$ by setting in the coordinates
above the coordinate expression of $\tilde{g}_{ab}$ equal to
\[
\beta(t,\mbox{{\bf x}}) dt^2 - (( 1 -f(t,\mbox{{\bf
    x}}))h_{ij}(t,\mbox{{\bf x}}) + f(t,\mbox{{\bf x}})
\tilde{h}_{ij}(t,\mbox{{\bf x}}) ) d\mbox{x}^i d\mbox{x}^j
\]
By choosing $\beta$ sufficiently small outside the region where it is
demanded to be one\footnote{%
I. g.\ we ``close'' up the light cone.},
we can ensure that $(\tilde{M}:={\Bbb R}\times \cC,\tilde{g}_{ab})$
is globally hyperbolic. Setting $U:= \phi^{-1}(\tilde{U})$ and
$\tilde{S} := \phi^{-1}(\hat{\cC}(t_1))$ finishes the construction.\\[0.5cm]
%%%%%%%%%%%%%%%%%%%%%%%%%%%%%%%%%%
%
A first consequence of Corollary~\ref{Cor:wavefrontscalar} is
\begin{Cor}\label{Cor:scalar_power}
  Let $\omega_n$ be the n-point distribution arising from a quasifree
  Hadamard state of a massive Klein-Gordon field propagating on a
  globally hyperbolic spacetime $(M,g_{ab})$. Then all finite powers
  of $\omega_n$ exist as products of distributions.
\end{Cor}
\begin{pf}
  Assuming all odd n-point distributions to vanish, we have since
  $\omega_n$ arises from a quasifree state
  \[
  \omega_n(x_1,\ldots,x_n) = \sum_P \prod_r \omega_2(x_{(r,1)},x_{(r,2)}),
  \]
  where $P$ denotes a partition of the set of points $\{ x_i\}$ into
  subsets which are pairings of points, labeled by $r$. Note that the
  ordering of the points in $\omega_2$ is preserved, e.g.\ $(r,1) <
  (r,2)$ and no two arguments are identical. The latter fact ensures
  the existence of the product $\prod_r$ whenever $\omega_2(x_i,x_j)$
  are distributions. For the wave front set of $\omega_n$ one finds
  using Theorem~\ref{Thm:IX.54}
  \begin{equation}
    \label{eq:wf_omega_n}
    \begin{split}
      &\WF(\omega_n) \\
      & = \bigcup_{(x_1,\ldots,x_n)\in M^n}
                        \WF\left(\omega_n(x_1,\ldots,x_n)\right) \\
                    & \subseteq \bigcup_{(x_1,\ldots,x_n)\in M^n}
                    \left(
                      \bigcup_P \bigcup_p \bigoplus_{r_p}
                      \left[ M^{n-2} \times
                        \WF\left( \omega_2(x_{(r_p,1)},x_{(r_p,2)})
                            \right)
                      \right]
                    \right),
    \end{split}
  \end{equation}
  where $p$ denotes a subset of $P$ and $r_p$ labels the elements of
  $p$.
  \begin{exmp} Consider the four-point distribution of a
    quasifree state:
    \begin{multline*}
      \omega_4(x_1,x_2,x_3,x_4) = \\
     \omega_2(x_1,x_2)\omega_2(x_3,x_4) +
     \omega_2(x_1,x_3)\omega_2(x_2,x_4) +
     \omega_2(x_1,x_4)\omega_2(x_2,x_3)
   \end{multline*}
   We have
    \[P \in \left\{ \{(x_1,x_2),(x_3,x_4) \},
    \{(x_1,x_3),(x_2,x_4) \}, \{(x_1,x_4),(x_2,x_3) \} \right\}
    \]
    Let
    $P= \{ (x_1,x_3), (x_2,x_4) \}$, then
    $p \in \left\{ \{(x_1,x_3)\}, \{(x_2,x_4)\}, \{
    (x_1,x_3),(x_2,x_4) \} \right\}$. Assume $p = \{ (x_1,x_3)\}$, then
    $r_p=(x_1,x_3)$, e.g.\ $x_{(r_p,1)}=x_1$ and
    $x_{(r_p,2)}=x_3$. Therefore
    \begin{equation*}
      \begin{split}
         &\WF(\omega_4)\\
        & \subseteq \bigcup_{(x_1,\ldots,x_4) \in M^4} \biggl(
        \left[ M^2 \times \WF\bigl(\omega_2(x_1,x_2)\bigr) \right] \cup
        \left[ M^2 \times \WF\bigl(\omega_2(x_3,x_4)\bigr) \right]\\
        & \phantom{\subseteq \bigcup_{(x_1,\ldots,x_4) \in M^4} \biggl(
        \left[ M^2 \times ) \right] \cup }
        \cup
        \left[
          \left( M^2 \times\WF\bigl(\omega_2(x_1,x_2)\bigr)
          \right)
          \oplus
          \left( M^2 \times\WF\bigl(\omega_2(x_3,x_4)\bigr)
          \right)
        \right] \\
        & \phantom{\subseteq \bigcup_{(x_1,\ldots,x_4) \in M^4}\biggl(}
        \cup 2 \leftrightarrow 3 \\
        & \phantom{\subseteq \bigcup_{(x_1,\ldots,x_4) \in
            M^4}\biggl(}
        \cup 2 \leftrightarrow 4 \biggr)\\
        & \equiv
        \left\{ (x_1,k_1;x_2,k_2;x_3,0;x_4,0) | \quad
          (x_1,k_1;x_2,k_2) \in \WF(\omega_2) \right\} \\
        & \phantom{\equiv} \cup
        \left\{ (x_1,0;x_2,0;x_3,k_3;x_4,k_4) | \quad
          (x_3,k_3;x_4,k_4) \in \WF(\omega_2) \right\} \\
        & \phantom{\equiv} \cup
        \left\{ (x_1,k_1;x_2,k_2;x_3,k_3;x_4,k_4) | \quad
          (x_1,k_1;x_2,k_2) \in \WF(\omega_2);\right. \\
        &  \left. \phantom{\equiv (x_1,k_1;x_2,k_2;x_3,k_3;x_4,k_4) | \quad
          (x_1,k_1;x_2,k_2)}
          (x_3,k_3;x_4,k_4) \in \WF(\omega_2) \right\} \\
        & \phantom{\equiv\quad} \cup \cdots
      \end{split}
    \end{equation*}
  \end{exmp}
  To prove the Corollary it is --by Theorem~\ref{Thm:IX.54}--
  necessary and sufficient to show that finite sums of $\WF(\omega_n)$
  do not contain zero. We know the wave front sets of all
  distributions in Eqn.~\ref{eq:wf_omega_n} explicitly (See
  Corollary~\ref{Cor:wavefrontscalar}). Consider the following two
  cases:
  \begin{enumerate}
  \item The direction in $\WF(\omega_n)$ associated to the first
    variable is not zero. Then its time component is strictly
    positive by Corollary~\ref{Cor:wavefrontscalar}. Since the time
    components of all directions associated to the first variable of
    $\WF(\omega_n)$ are always greater or equal to zero, the sum of
    all these directions can not vanish and the Corollary is proved.
  \item The direction in $\WF(\omega_n)$ associated to the first
    variable is zero for all summands. Then all directions associated
    to the second variable must have time components greater or equal
    to zero and we apply the same argumentation as for the first
    variable. Note that it is excluded by the definition of the wave
    front set that all directions of all variables vanish
    simultaneously. Therefore there exists a variable $(x_i)$ $i<n$
    such that the analogon of (1) holds.
  \end{enumerate}
\end{pf}
\subsection{A local characterization of globally Hadamard states}
One of the main motivations for using wave front sets in quantum field
theory is the fact that they allow the specification of global
properties locally. The following Theorem, which is one of
the main results of Radzikowski's dissertation, gives a local
characterization of globally Hadamard states.
\begin{Thm}[Theorem 2.6 of \cite{Radzikowski:92}]
  \label{Thm:locHadam}
  Let $\omega_2$ be the two-point distribution arising from a state of
  a massive Klein-Gordon field propagating on a globally hyperbolic
  spacetime. If $\omega_2$ has wave front set as in
  Corollary~\ref{Cor:wavefrontscalar} above then $\omega_2$ is
  globally Hadamard.
\end{Thm}
\begin{rem}
  Our assumption implies that $\omega_2$ fulfils the Klein-Gordon
  equation and has $i$ times the commutator distribution of $\Box +
  m^2$ as its antisymmetric part. The
  Corollary~\ref{Cor:wavefrontscalar} above may be viewed as the
  converse of this Theorem.
\end{rem}
The proof of this Theorem is rather long; we refer the reader to the
dissertation of Radzikowski for the details.  Radzikowski's
Theorem~2.6 states the equivalence of the wave front set assumption
and the globally Hadamard condition $\mod \Cinf$. To prove it, he
introduces the Feynman two-point distribution $\omega_F$ defined by
$\omega_F:= i \omega_2 + E^+$, where $E^+$ is the advanced fundamental
solution of $\Box + m^2$. The assumption on the wave front set of
$\omega_2$ uniquely fixes this distribution (up to $\Cinf$) to be the
Feynman two-point distribution defined in~\cite{Hoermander:72}.  This
in turn implies that $\omega_2$ is globally Hadamard $\mod
\Cinf$.
\message{Section: \thesection}
\section{The wave front set spectrum condition}
We have seen in the last section that the wave front set of a state
can be used to characterize this state globally. However
Equation~(\ref{eq:wfhadamscalar}) restricts the singular support of
$\omega_2(x_1,x_2)$ to points $x_1$ and $x_2$ which are null related;
hence $\omega_2$ is smooth for time-like or space-like related points.
The latter smoothness is known to be true for reasonable quantum field
theories on Minkowski space satisfying the true spectrum condition by
the Bargmann-Hall-Wightman Theorem. For time-like related points
however a similar general prediction on the smoothness does not exist.
In order to include possible singularities at time like related
points, Radzikowski extended in~\cite{Radzikowski:92} the right hand
side of Eqn.~(\ref{eq:wfhadamscalar}) to all causally related points;
he proposed that the wave front set of the two-point distributions of
any physical reasonable state should be contained in this extended
set. He called this proposal the `wave front set spectrum condition'
(WFSSC).  He also proposed a WFSSC for higher n-point distributions
and showed that both of his Definitions are compatible to the usual
spectrum condition: Each $\omega_n$ fulfiling his WFSSC satisfies the
true spectrum condition on Minkowski space $\mod \Cinf$ (Theorem~4.10
of~\cite{Radzikowski:92}) and vice versa. He gives further evidence on
the legitimacy of his Definitions by linking them to the scaling limit
condition of Fredenhagen and Haag~\cite{fredenhagen:87}: Both
Definitions imply the true spectrum condition in the scaling limit if
this limit exists (Theorem~4.11 of~\cite{Radzikowski:92}).
Unfortunately it can be shown that the n-point distributions for $n >
2$ associated to a quasifree Hadamard state of a scalar field on a
globally hyperbolic spacetime do not satisfy his WFSSC in general.
Thus the WFSSC, at least for the higher n-point distributions, needs
to be modified. We do not know a reasonable modification at the moment
and therefore restrict ourself in this note to the WFSSC for the
two-point distributions.

The counterexample mentioned above and the wish to include fields
which are a composition of `simple' fields (such as our field $K$),
leads us to propose the following `conic'
WFSSC\footnote{E.\ g.\ $(x,k_1;y,l_1),(x,k_2,y,l_2) \in \WF(\omega_2)
  \Rightarrow (x,\lambda k_1 + \mu k_2; y, \lambda l_1 + \mu l_2) \in
  \WF(\omega_2) \quad \forall \lambda,\mu > 0$.} for the two-point
distributions, which is a slight modification of Radzikowski's original
definition.
\begin{Dfn}[WFSSC]\label{dfn:wfspectrum}
  The two-point distribution $\omega_2 \in {\cal D}'({M\times M})$
  satisfies the {\em wave front set spectrum condition (WFSSC)\/} iff
  its wave front set $\WF(\omega_2)$ consists only of points
  $(x_1,k_1),(x_2,k_2) \in T^*M \setminus \{ 0 \}$ such that $x_1$ and
  $x_2$ are causally related and $k_1$ is in the closed forward
  lightcone.  Furthermore there are causal geodesics $\gamma_i$
  joining $x_1$ and $x_2$ and vectors $k_i$ in the closed forward
  lightcone, such that $\sum_i k_i =k_1$ and the parallel transported
  vectors $k_i$ along $\gamma_i$ sum up to $-k_2$.\footnote{Note that
    our definition allows the singularities to propagate along
    multiple curves $\gamma_i$ simultaneously.}
\end{Dfn}
%% %%
\section{Proof of Proposition~\ref{prop:strom}}
It is first shown that $K$ is well defined and fulfils the following
four Wightman axioms:
\begin{enumerate}
\item The distributional product of $A$ and $B$, $K$, is a well
  defined operator valued distribution on the GNS Hilbertspace
  $(H,\Omega,A,B)$ with domain
  \[ D= \mbox{Span} \{ \Phi(f_1) \cdots \Phi(f_n) |\Omega> \, | \quad
  \Phi \in \{A,B,K\},
  f_1,\ldots,f_n \in \CinfO(M)
  \}
  \]
\item $D$ is dense in $H$.
\item For all $f \in \CinfO (M)$, $K(f)$ leaves $D$ invariant.
\item For all $f \in \CinfO (M)$, $K(f)$ is local (e.g.\ bosonic).
\end{enumerate}
We finish the proof showing that the two-point distribution of $K$
with respect to the ``vacuum'' $\Omega$ fulfils the
WFSSC~(Definition~\ref{dfn:wfspectrum}).
\begin{enumerate}
\item Let $D'= \mbox{Span} \{ \Psi = K(f_1) \cdots K(f_n) |\Omega>\, |
  \quad f_i
  \in \CinfO(M) \}$. We remark that the definition of $D'$ is only
  formal at this stage. To show that $K$ is well
  defined on $D'$, it is sufficient to show that
  \begin{alignat}{2}
    \left\| K(f) \Psi \right\| & < \infty && \qquad \forall f \in
      \CinfO(M), \quad \Psi \in D' \label{norm}
  \end{alignat}
  Inserting the definitions of $K$ and $\Psi$ in Equation~(\ref{norm})
  we obtain in local coordinates:
  \begin{equation}  \label{eq:normkf}
    \begin{split}
      {\| K(f) \Psi \|}^2
      & = < K(f)K(f_1)\cdots K(f_n) \Omega | K(f)K(f_1)\cdots K(f_n)\Omega> \\
      & = <\Omega |
        K(\bar{f}_n)\cdots K(\bar{f}_1)K(\bar{f})K(f)K(f_1)\cdots
        K(f_n)\Omega> \\
      & = \idotsint d\mu_{x_n} \cdots d\mu_{x_1} d\mu_x d\mu_y
                  d\mu_{y_1} \cdots d\mu_{y_n} \\
      & \phantom{=}
        \sideset{^A}{_{2n+2}}{\omega}
                (x_n,\ldots,x_1,x,y,y_1,\ldots,y_n) \cdot \\
      & \phantom{=\sideset{^A}{_{2n+2}}{\omega}
                (x_n,\ldots,x_1)}
        \sideset{^B}{_{2n+2}}{\omega}
                (x_n,\ldots,x_1,x,y,y_1,\ldots,y_n)\\
      & \phantom{=}
        \bar{f}_n(x_n) \cdots \bar{f}_1(x_1) \bar{f}(x) f(y) f_1(y_1)
        \cdots f_n(y_n)
    \end{split}
  \end{equation}
  Equation~(\ref{eq:normkf}) separates into the distributional product
  of two $2n+2$-point distributions since we are dealing with a
  product state.  $\sideset{^\cdot}{_{2n+2}}{\omega}$ denotes the
  $2n+2$-point distribution of the corresponding basic field $A$ or
  $B$ and the r.h.s.\ of Eqn.~(\ref{eq:normkf}) {\em defines\/} $\|
  K(f) \Psi \|^2$ . All $2n+2$-point distributions arise from
  quasifree states.  Hence they decay into the sum of products of
  two-point distributions. However using
  Corollary~\ref{Cor:scalar_power} one sees that arbitrary (finite)
  products of these distributions with each other exist and are again
  distributions. This ensures the finiteness of
  Eqn.~(\ref{eq:normkf}). Now suppose we replace one or more $K(f_i)$
  in Eqn.~(\ref{eq:normkf}) by
  $\phi(g_1,\ldots,g_k,\tilde{g}_1,\ldots,\tilde{g_l}) \equiv A(g_1)
  \cdots A(g_k) B(\tilde{g}_1) \cdots B(\tilde{g}_l) $ with
  $g_1,\ldots,\tilde{g_l} \in \CinfO (M)$.  Redoing the calculation in
  Eqn.~(\ref{eq:normkf}) one ends up again with various products of
  two-point distributions. However all two-point distributions which
  include parts of the ``$\phi$'s'' are in fact smooth functions (for
  example $\sideset{^A}{_2}{\omega}(g_i,x) \in \Cinf(M);
  \sideset{^A}{_2}{\omega}(g_i,g_j) \in {\Bbb C} $). Thus replacing
  an operator $K(f_i)$ by $\phi(g_1,\ldots,\tilde{g}_l)$ does not
  change the finiteness of the whole equation. Furthermore, since the
  products in Eqn.~(\ref{eq:normkf}) are again {\em distributions}, it
  follows that $\tilde{K}_{\Psi \Psi'}: \CinfO (M) \ni f \mapsto <\Psi
  | K(f) \Psi' >$ is well defined and continuous for all $\Psi, \Psi'
  \in D$. This proves the assertion.
\item is obvious since $H$ is the GNS Hilbertspace of the basic
  fields, e.g.\ the ``vacuum'' $\Omega$ is cyclic for $A$ and $B$.
  \item follows directly from our definition of $D$.
\item To prove 4.\ one notes that $A$ and $B$ mutually commute and
  \[ [A(f),A(g)]_- = [B(f),B(g)]_- = 0 \qquad
                     \text{if $\supp(f) \subset \supp(g)^c$},
  \]
  for causally disjoint arguments.
  $\supp(g)^c$ denotes the causal complement of the support of $g$. Therefore
  \[ [ K(f),K(g) ]_- = 0 \qquad
                    \text{if $\supp(f) \subset \supp(g)^c$},
  \]
\end{enumerate}
\begin{rem}
  \begin{itemize}
  \item $D'$ is not dense in $H$ since for example $\Phi = A(f)
    |\Omega>$ can not be approximated by elements in $D'$: $\forall D'
    \ni \Psi$ we have $\| \Phi - \Psi \|^2 = \| \Phi\|^2 + \|\Psi\|^2$
    since $<\Phi,\Psi> = 0$. The latter equation is valid, since
    $<\Phi,\Psi>$ decays into a product of n-point distributions, one
    of which has an even, the other an odd number of arguments; such
    combinations always vanish.
  \item The vector $\Omega$ is not cyclic for the field $K$.
  \end{itemize}
\end{rem}
To show that the two-point distribution of $K$ in $\omega$ fulfils the
WFSSC, we calculate its wave front set.  In local coordinates, we
obtain:
\begin{equation}\label{eq:omega_2}
 \begin{split}
   \sideset{^K}{_2}\omega(f,g)
   & := < \Omega | K(f) K(g) \Omega> \\
   & = \underset{M \times M}{\iint} d\mu_x d\mu_y
       \sideset{^A}{_2}{\omega}(x,y) \sideset{^B}{_2}{\omega}(x,y)
       f(x) g(y)
 \end{split}
\end{equation}
Equation~(\ref{eq:omega_2}) is just the product of distributions
on the manifold $M$ and
by Corollary~\ref{Cor:wavefrontscalar} above
\begin{multline*}
\WF(\sideset{^A}{_2}\omega) = \WF(\sideset{^B}{_2}\omega) = \{ (x,k) ,
(y,k') \in (T^*M \times T^*M) \setminus \{ 0 \} | \\
 (x,k) \sim (y,- k'), k^0 > 0 \}.
\end{multline*}
Since $\WF(\sideset{^A}{_2}{\omega}) \oplus
\WF(\sideset{^A}{_2}{\omega})$ fulfils the assumption of
Theorem~\ref{Thm:IX.54}, we deduce that the kernel of
(\ref{eq:omega_2}) defines a unique distribution, namely
$\sideset{^K}{_2}{\omega}$ with wave front set contained in
\[
W:= \left(\WF(\sideset{^A}{_2}\omega) \oplus
\WF(\sideset{^A}{_2}\omega)\right) \cup \WF(\sideset{^A}{_2}\omega )
\]
We finish the proof showing that the elements of $W$ fulfil the assumption of
Definition~\ref{dfn:wfspectrum} above: \\
Consider an element $(x,k;y,l)$ of $W$. If it is contained in
$\WF(\sideset{^A}{_2}\omega)$, everything is proved, since
$\WF(\sideset{^A}{_2}\omega)$ is assumed to be the two-point
distribution of a Hadamard state. On the other hand if this element is
contained in $\WF(\sideset{^A}{_2}\omega) \oplus
\WF(\sideset{^A}{_2}\omega)$, then there exist covectors $k_1$, $k_2$
and $l_1$, $l_2$ respectively with $k_1 + k_2=k$ and $l_1 + l_2 = l$,
such that $(x,k_i;y,l_i) \in \WF(\sideset{^A}{_2}\omega)$ for $i=1,2$.
With the corresponding curves $\gamma_i$ we have
\begin{enumerate}
\item $k_1 + k_2 \in V_{+x}$ since $k_i \in V_{+x}$ by assumption.
\item $g^{\gamma_1}_{12}k_1 + g^{\gamma_2}_{12} k_2 = - (l_1 + l_2)$
  since $(x,k_i) \sim (y,- l_i)$
\end{enumerate}
$V_{+x}$ is the closed forward lightcone at $x \in M$ and
$g^{\gamma_i}_{12}$ denotes the parallel transport along the curve
$\gamma_i$ from $x_1$ to $x_2$.
\begin{rem}
  \begin{itemize}
  \item In general there exists more than one causal curve
    connecting $x_1$ and $x_2$. The modification in
    Definition~\ref{dfn:wfspectrum} which allows multiple curves
    $\gamma$ to contribute is necessary because in the generic case
    the parallel transport is path dependent, e.g.\
     \[
     g_{12}^{\gamma_1} k_1 + g_{12}^{\gamma_1} k_1 \neq - (l_1 + l_2)
     \neq g_{12}^{\gamma_2} k_1 + g_{12}^{\gamma_2} k_1
     \]
   \item The two-point distribution of our state is smooth for
     time-like related points. It is therefore possible to strengthen
     Definition~\ref{dfn:wfspectrum} by demanding that the singular
     support of $\omega_2$ should contain light-like related points
     only.
  \end{itemize}
\end{rem}
\section{Summary and Outlook}
In this note it has been proved that the distributional product of two
scalar fields on a globally hyperbolic spacetime gives a new Wightman
field on this manifold. Obviously the result extends to all cases
where the definition of a quasifree Hadamard state makes sense. The
free Dirac field is an example, which will be used in a forthcoming
paper to calculate the renormalized stress energy tensor of an
analogon of the free Wess-Zumino model on a manifold.  Moreover it has
been shown that this Wightman field satisfies the new wave front set
spectrum condition (WFSSC) on a manifold. In the opinion of the author
this condition will turn out to be a valuable tool for quantum field
theory on curved spacetimes. It should for example enable some kind of
perturbation theory on manifolds in the future. But even on Minkowski
spacetime the study of wave front sets of more realistic models (for
example QED) might result in a deeper understanding of the Wightman
axioms and the true spectrum condition.

An extension of the WFSSC to higher $n$-point distributions remains to
be found, since the proposal of Radzikowski is not acceptable.
\begin{ack}
The author has benefited from discussions with many members of the
II.~Institut f\"ur Theoretische Physik at the University of Hamburg.
In particular he wishes to thank Prof.~Fredenhagen for his guidance
during this research, for many stimulating discussions and for
valuable comments on the manuscript. Special thanks are due to R.~Verch
for helpful hints and a careful reading of the manuscript.
M.~Radzikowski is gratefully acknowledged for having made the results of
his PhD.~thesis available prior to publication.
\end{ack}
%
%% Bibliography
%\bibliographystyle{amsalpha}
%\bibliographystyle{amsplain}
% amsalpha oder alpha
%\bibliography{bibspin}
\makeatletter \renewcommand{\@biblabel}[1]{\hfill#1.}\makeatother
\end{document}